\documentclass[10pt,journal,compsoc]{IEEEtran}
\usepackage{tabularx}
\usepackage{url}

\title{Network Security Roadmap}

\author{Praveen Kumar}

\date{}

\usepackage[numbers]{natbib}
\usepackage{graphicx}

\begin{document}

\maketitle

\section{Abstract}
Users may already have some perception of provided security based on experience with earlier generations. To maintain the stability and coherent integration of 5G services, it is imperative that security and privacy features prevalent in earlier generations are also present in 5G. However, it is not sufficient just to provide the same security features as in the legacy systems due to the new threat model introduced by the integration of new technologies like SDN, virtualization and SBA. 5G systems are expected to be more service-oriented. This suggests there will be an additional
emphasis on security and privacy requirements that spawn from the new dimension of service-oriented security architecture.

\section{Introduction}

The fifth generation network, popularly referred to as 5G, has the potential of enabling multiple vertical industries~\cite{vannithamby2017towards, saha2021sharks} like vehicle networks~\cite{chen2017vehicle, molina2017lte}, IoT~\cite{mavromoustakis2016internet, saha2021sharks}, VR/AR~\cite{erol2018caching}, smart cities~\cite{skouby2014smart, santos2018fog, brown2021gravitas} and smart healthcare~\cite{chen20185g,chen20175g,latif20175g}. The millimeter wave technology that 5G employs offers 100x lower latency and 100x more bandwidth~\cite{chavez20155g} than the existing 4G technology. As a result of this, real-time security requirement is a prime concern for the deployment of 5G applications. When it comes to networks involving the Internet of Things (IoT), multiple layers of the stack are vulnerable as potential targets of security attacks~\cite{ahmad20175g}, including the service and application layer, nodes, platforms, and the network/transport layer. For example, a unified framework of vulnerability detection in an IoT system is described in ~\cite{brown2021gravitas}. The increased speed and bandwidth enabled by 5G will eventually lead to new threat vectors and increased sophistication of cyber attacks.

Another growing concern is the acceleration of existing cyber attacks in the 5G framework. Today's cyber attacks have already targeted networks' security: both at the IP layer and signalling layer. Thus, simply making legacy security mechanisms operate faster is not an adequate measure. The Service Based Architecture (SBA) of the core 5G network provides multiple access points to third-party service providers, thus expanding the attack surface. Legacy approaches that depend on disparate security elements will neither scale nor will be able to adequately prevent successful attacks across the SBA of 5G networks~\cite{ferrag2018security}. New encryption schemes and authentication protocols like 5G AKA~\cite{cremers2019component} are being designed for this purpose. System designs composed of individually secure subsystems are not provably secure. 5G radio network deployments include substantial augmentations of small cells connecting over insecure networks, edge communications, and edge nodes communicating with multiple cells concurrently~\cite{bangerter2014networks}. Moreover, 5G comes with its new authentication protocols and encryption primitives, which have not received a sufficient amount of cryptanalysts' scrutiny yet. This evolution broadens the threat landscape by expanding the number of intrusion points. With billions of connected devices and critical enterprise applications relying on 5G networks, network operators and applications should preemptively defend against possible security breaches. Some effective measures for ensuring comprehensive end-to-end (E2E) security in the 5G framework may include:

\begin{itemize}
    \item Multiple factors of mutual authentication across all layers of the network including application, control and data planes~\cite{ni2018efficient, ferrag2018security, dubrova2015crc}.
    
    \item Cloud-based threat analysis and vulnerability management, enabled by machine learning(ML) algorithms, can be used across multiple platforms to provide immediate response to known and unknown threats in real time~\cite{zhang2016scanme, yu2013cloud, chen2016cloud}.
    
    \item The explosion in the number of end devices requires real-time quarantining of infected devices using data-driven approaches.
    
    \item Multi-layered encryption of software-defined network (SDN) data planes, enabled by data classification is required for confidentiality and integrity protection.
\end{itemize}

\begin{figure*}[h!]
\centering
\includegraphics[scale=0.3]{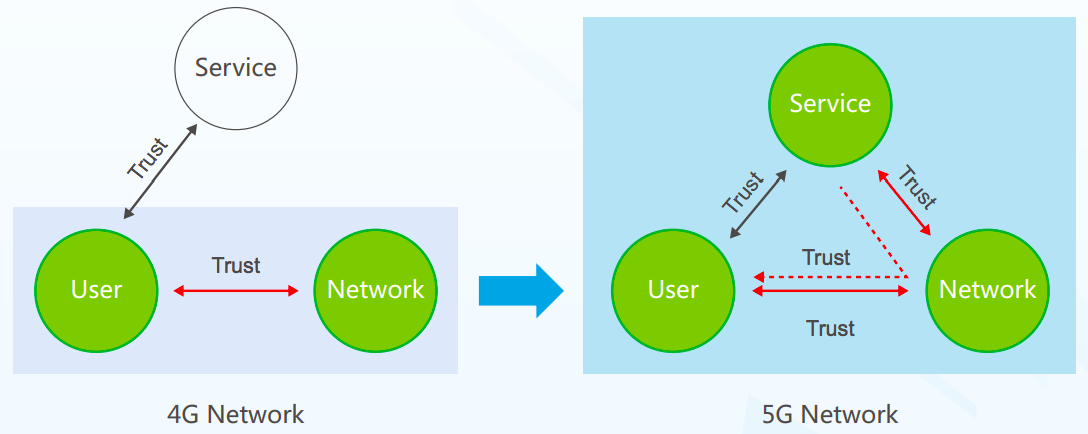}
\caption{Evolved trust model in 5G~\cite{Web1}}
\label{fig:trust_model}
\end{figure*}

A primary reason for the necessity of stricter security measures in emerging networks is the change of the trust model as depicted in Fig.~\ref{fig:trust_model}.

Security for 5G should be taken into consideration at the initial stages of the system design (Security by Design approach). Augmenting security features at a later stage is less efficient and often more expensive in the long run. In long term, security is a driving factor for service and network evolution.

\section{Background}
We provide a brief background on relevant prior work in this field.

\subsection{5G Network Security}

Unlike previous generations of communication, the security vulnerabilities in the 5G infrastructure are manifold. This is due to the integration of multiple 5G enabling technologies that alter the traditional architecture of 4G networks~\cite{bangerter2014networks, chen2014requirements}. The enhanced mobile broadband and high reliability come with the introduction of flexible mobility and networks. Programmable networks, network function virtualization (NFV) and network slicing are some of the technologies that make the 5G dream possible~\cite{boccardi2014five, wang2014cellular}. However, the integration of these features increase the attack surface. The SBA of 5G has multiple third-party service providers sharing resources for efficient functioning of the network. This makes every entity a potential point of attack, unlike the existing systems where most services are provided by the network provider. Hence, 5G security is not just an extension of the current security measures.

The primary changes in the 5G Core from the existing 4G Evolved Packet Core manifest in the form of NFV, Multi-access Edge Computing (MEC)~\cite{kekki2018mec,hu2015mobile}, Radio Access Network (RAN)~\cite{parvez2018survey} and SDN~\cite{blanco2017technology, trivisonno2015sdn, cho2014integration}. NFV decouples software from hardware by replacing various network functions such as firewalls, load balancers and routers with virtualized instances. Network slicing enables the implementation of NFV by allowing multiple logical networks to run on top of a shared physical resource. This provides increased flexibility across networks. SDN is a complimentary technology to NFV for achieving a unified network abstraction to expand data flow control in the form of virtual switches~\cite{ordonez2017network}.

\section{Proposed Work}
There have been preliminary studies on the emerging vulnerabilities in the individual components of the 5G network~\cite{geller20185g, ahmad20175g, yan2016security, vassilakis2017security, hussein2017sdn}. However, there has not been a comprehensive analysis of the vulnerabilities emerging due to integration of these various components yet. We aim to study such vulnerabilities. There are some attack classes that can be more effectively defended against in the 5G framework. We wish to study such scenarios as well. We also aim to study system level (OS and kernel) and application level vulnerabilities, their correlation to network vulnerabilities as well as the emergence of new threats using artificial intelligence methods. 

\subsection{Pillars of the methodology}
Newer technologies integrated in 5G require the formulation of:
\begin{itemize}
    \item \textbf{New stream ciphers:} As mentioned earlier, the 5G ecosystem will enable multiple vertical industries like IoT. This involves the existence of a heterogeneous amalgamation of resource-constrained edge devices. More energy-efficient encryption algorithms and authentication methods are required for efficiency of these devices. Some good candidates for efficient 5G radio communications with constrained resources are mentioned in ~\cite{Dubrova2017, dubrova2015crc}. However, in-depth cryptanalysis is yet to demonstrate the strength of such ciphers. 
    
    \item \textbf{New security protocols:} Due to the change in the architecture of existing 4G networks, new protocols have been designed for better attack resistance of 5G radio access networks. Random access procedure based on tunable puzzles is an example introducing generalization and improved usability in existing security puzzles~\cite{saha2016tv, sehwag2016tv}. 
    
\end{itemize}

On another hand, 5G network is composed of many subsystems functioning together. The security of each of these subsystems as stand-alone disparate systems is necessary but not sufficient to ensure the security of the overall system. The study of the security challenges arising due to the integration of the following subsystems is still in its infancy. Some of the major 5G security subsystems are:
\begin{enumerate}
    \item \textbf{Application layer security}: Application layer security refers to ways of protecting the application layer from malicious attacks and the exploit of inherent software vulnerabilities. Application layer provides the largest threat surface to the attackers due to accessibility and the huge diversity and minimal abstraction from the user. Application layer vulnerabilities may lead to performance and stability degradation, loss of privacy, and the network being taken down~\cite{zhang2017overview}. Application layer threats include buffer overflows, race conditions, SQL injections and privilege escalation, amongst others. 
    
    \item \textbf{Edge security:} The edge of the network generally comprises the computing nodes, like smartphones, sensors, RFID tags and smart controllers. Edge devices often process and store confidential information which must be protected. In a smart system like an autonomous vehicle or an IoT ecosystem ~\cite{fan2016lightweight, mavromoustakis2016internet}, an edge device (namely the microcode, kernel and operating system layers) are prone to security exploits. It is essential to detect such exploits in real-time. Edge security poses  many different threats, a major one being a DDoS attack launched by hijacked edge nodes. 
    
    \item \textbf{Cloud RAN (C-RAN) security:} C-RAN is a centralized cloud-based radio access network architecture that has multiple advantages over traditional radio access networks. C-RAN has been adopted by multiple operators for the 5G infrastructure~\cite{wubben2014benefits}. Due to its widespread integration, C-RAN security has received a lot of attention recently. The traditional C-RAN threat vectors include eavesdropping attacks, impersonation attacks, jamming attacks, primary user emulation attacks and wireless channel attacks. Most of these attacks can be evaded using existing cybersecurity measures but the potential emergence of novel threat vectors due to its integration to the disparate new technologies have not been explored yet~\cite{parker2016charisma, demestichas20135g}.
    
    \item \textbf{SDN security:} Traditionally, network components like routers and switches have the rules for packet forwarding hard-coded into the devices. This inhibits the flexibility of the devices and may lead to inefficient packet routes. SDN separates the control plane and the data plane, allowing a software defined controller to dynamically determine the packet forwarding routes. This increases efficiency. Many 5G operators are looking forward to integrate SDN in the 5G framework. However, the SDN controllers are stand-alone components which need to be protected from attackers~\cite{duan2015authentication, duan2016fast}. Other security aspects includes establishing trust between SDN controllers and the data plane (via authentication), creating a robust policy framework and conducting forensic tests to detect and remediate a compromised controller. Various attacks include DoS attacks on the controller and the data planes, topology poisoning attacks and man-in-the-middle (MiTM) attacks. These attacks are described in detail in section~\ref{SDN_attacks}.
    
    \item \textbf{Proactive security analysis:} The 5G infrastructure is a service based architecture. Such an architecture is controlled by smart controllers. Hence, future updates in 5G will be software updates. Because of the cyber vulnerabilities of software, we need a proactive real-time security analyzer to detect and alleviate the threats before an attacker exploits it~\cite{perez2017dynamic, siddiqui2016policy}.
    
    The dramatic increase in bandwidth in 5G gives rise to additional attack vectors. Short range, small-cell antennae physically deployed all across the country become the new entry-points for attackers, expanding the attack surface. Functionally, these antennae will use 5G’s Dynamic Spectrum Sharing capability in which multiple streams of information share the bandwidth in so-called “slices”—each slice with its own varying degree of cyber risk. Such dynamic modes of operation require dynamic cybersecurity solutions, rather than uniform rigid defences.

    \item \textbf{Hypervisor security:} Hypervisors enable the mapping of various network functions to underlying logical instances and hardware in a virtualized 5G network. A hypervisor can create and execute multiple guest operating systems and performs the necessary resource allocation for those systems. This makes the hypervisor a potential point of entry to multiple networks. Hypervisors are susceptible to DoS attacks on virtual machines (VMs), VM hopping attacks and network slice attacks via host operating systems. Some common hypervisor vulnerabilities include software bugs and network attacks like network intrusion and DoS attacks.
    
    \item \textbf{Faster authentication:} With the facilitation of multiple vertical industries like autonomous vehicles, IoT and smart healthcare, the number of edge devices will face an exponential growth. The number of virtual machines will also increase due to the integration of services like SDN and NFV. All these individual components need to authenticate themselves to one another to ensure secure interaction between them~\cite{norrman2016protecting, basin2018formal}. Faster and reliable authentication mechanisms and policies are required for effective operation of the 5G infrastructure.
    
    \item \textbf{NFV security:} NFV utilizes virtualization to allow multiple networks to run on shared resources. NFV is an integral part of the 5G ecosystem and securing it is a fundamental concern. While some of the issues related to NFV security like hypervisor security and SDN security have been discussed before, there are certain unique threats to NFV which must be considered during secure 5G system design ~\cite{lal2017nfv, yang2016survey, bernardo2015introduction, shih2016s}. These vulnerabilities include software bugs, insider attacks (like Trojans), outsider attacks (like malicious third-party operator) and attacks on shared resources (like cache poisoning attacks)~\cite{aljuhani2017virtualized}.
    
    An advantage of NFV is that it allows the virtualization of network security functions like intrusion detection systems. This enables security-as-a-service infrastructure, which is integral to the service based architecture of 5G.
    
\end{enumerate}

\subsection{Preliminary Evaluation}

\subsubsection{Results on WhatsApp}
We consider WhatsApp as our secure chat, voice and video application running on a 5G network for our security evaluations. We propose to apply cyber-analysis to WhatsApp as a standalone application which implements several in-transit and at-rest security countermeasures. We will also analyze WhatsApp in the context of 5G networks. 

WhatsApp is a secure instant messaging application with over 1.5 billion users in over 180 countries. The WhatsApp messenger uses XMPP protocol and enforces E2E encryption. E2E encryption claims that no one other than the end users can decrypt messages in transit (including proxy and authentication servers). WhatApp E2E encryption is built on top of the security protocol used by the highly secure messaging application Signal. However, there are certain security measures that have been omitted by WhatsApp for higher efficiency. As a stand-alone application, WhatsApp is highly secure. Morever, security vulnerabilities at different layers like the network, IP or transport layer can be exploited to compromise the security of WhatsApp. We aggregate all the known attack vectors on WhatsApp across the application layer, network layer, authentication interfaces and the edge device~\cite{saha2022machine, saha2022machine5g}. Till now, we have constructed a DAG of 5 classes of vulnerabilities. The DAG consists of 23 nodes currently. We will keep on augmenting our DAG as new vulnerabilities are exposed. The various threat vectors are:

\begin{enumerate}
    \item \textbf{Image file-jacking:} This attack vector is launched by a malware or rootkit on the smartphone which has access to the shared memory. Sandboxing allows applications like WhatsApp to store its messages in a secure partition of the memory which is inaccessible to any other third-party application. However, the images and other media contents received via the messenger are stored in the shared external memory (/storage/emulated/0/Whatsapp/media). There is usually a delay between receiving the media file in memory and loading it in the chat user interface. The malware can potentially alter the media file during this time and the attack will go undetected. This is also known as the Man-in-the-Disk attack and can be used to manipulate images, payment information and audio files. The Control Flow Graph (CFG) of this attack is shown in Fig.~\ref{fig:Image_file_jacking}.
    \begin{figure*}[h!]
    \centering
    \includegraphics[scale=0.7]{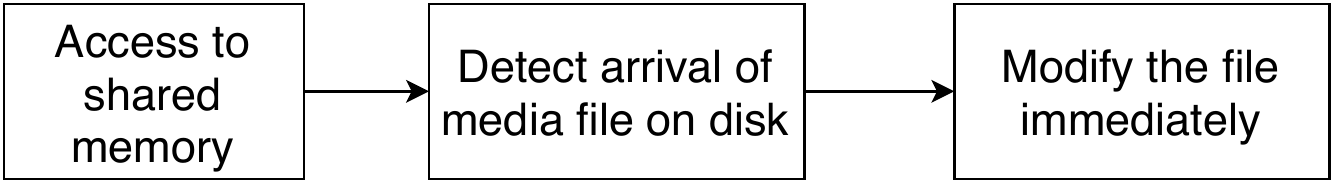}
    \caption{CFG of image file jacking}
    \label{fig:Image_file_jacking}
    \end{figure*}
    
    \item \textbf{Insecure storage of keys:} The WhatsApp encryption keys are stored on the device in clear text. It is assumed that sandboxing prevents any other application from accessing it~\cite{saha2022machine5g}. However, root access to the phone and memory corruption attacks, enables access to key chains and renders encryption keys vulnerable to confidentiality and integrity attacks respectively. Besides these, there are methods to decrypt messages on an older smartphone after migrating to an existing WhatsApp account on a new device. These methods can be exploited to extract the private key.

    \item \textbf{SS7 network protocol exploit:} WhatsApp verifies a user's identity with a one-factor authentication, namely the phone number used for registration, by calling the user's registered phone number or by sending him/her a verification code via SMS. The network protocol Signalling System 7 (SS7), which is used for communication by a large number of providers, is vulnerable and prone to session hijacking. A malicious adversary can hijack the authentication call (or SMS) and verify himself as the user, thus impersonating the user in future conversations. The CFG of this attack is shown in Fig.~\ref{fig:SS7}.
    \begin{figure*}[h!]
    \centering
    \includegraphics[scale=0.7]{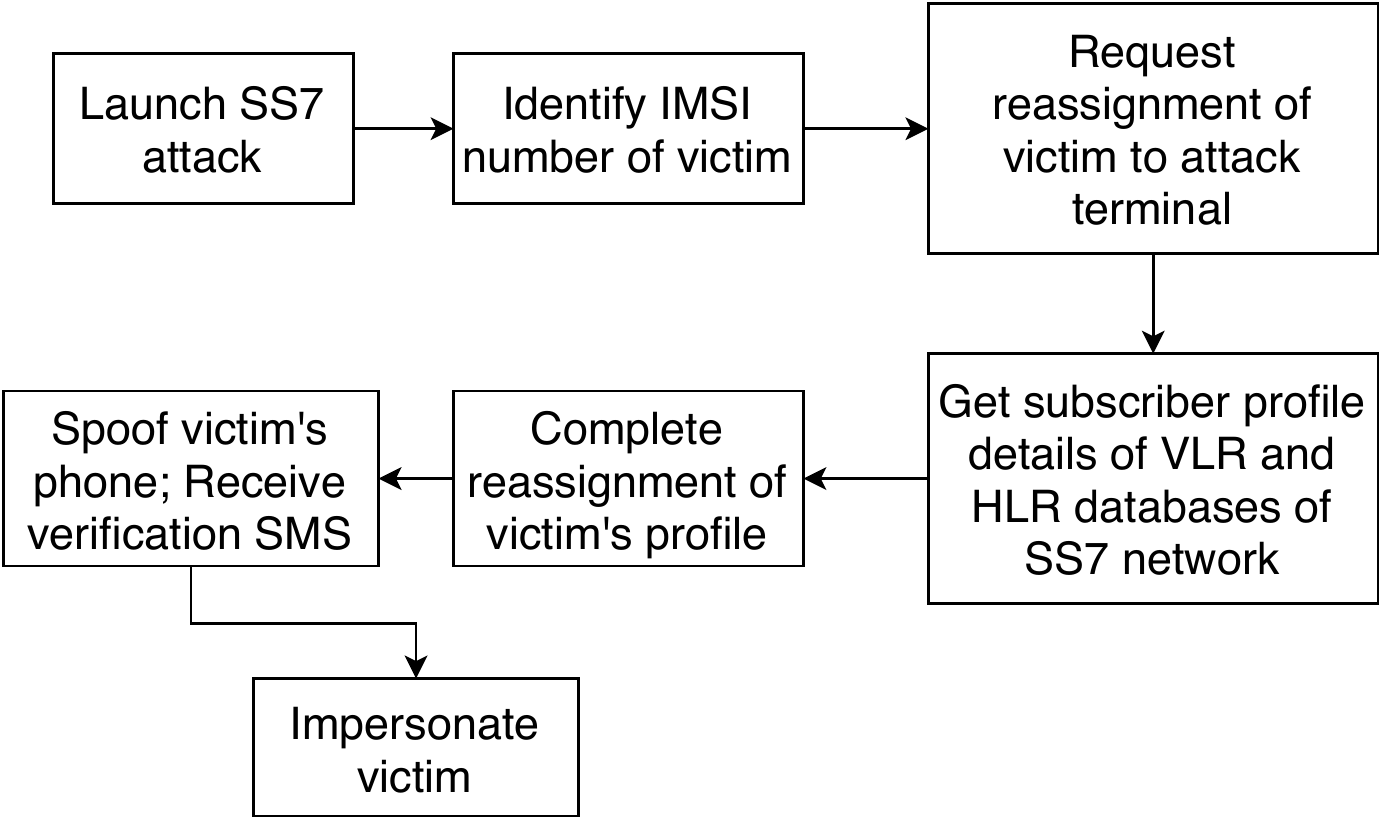}
    \caption{CFG of SS7 vulnerability exploit}
    \label{fig:SS7}
    \end{figure*}
    
    \item \textbf{Voicemail exploit:} As mentioned before, WhatsApp uses voice calls to verify a new account. If the user doesn't receive the verification call, the verification code is sent to the user's voicemail. Voicemail accounts can be hacked easily with brute-force attacks on voicemail account passcodes. Hacking the voicemail gives access to the verification code which can be exploited by the attacker to impersonate the user in future conversations. The CFG of the voicemail exploit is depicted in Fig.~\ref{fig:Voicemail}.
    \begin{figure*}[h!]
    \centering
    \includegraphics[scale=0.6]{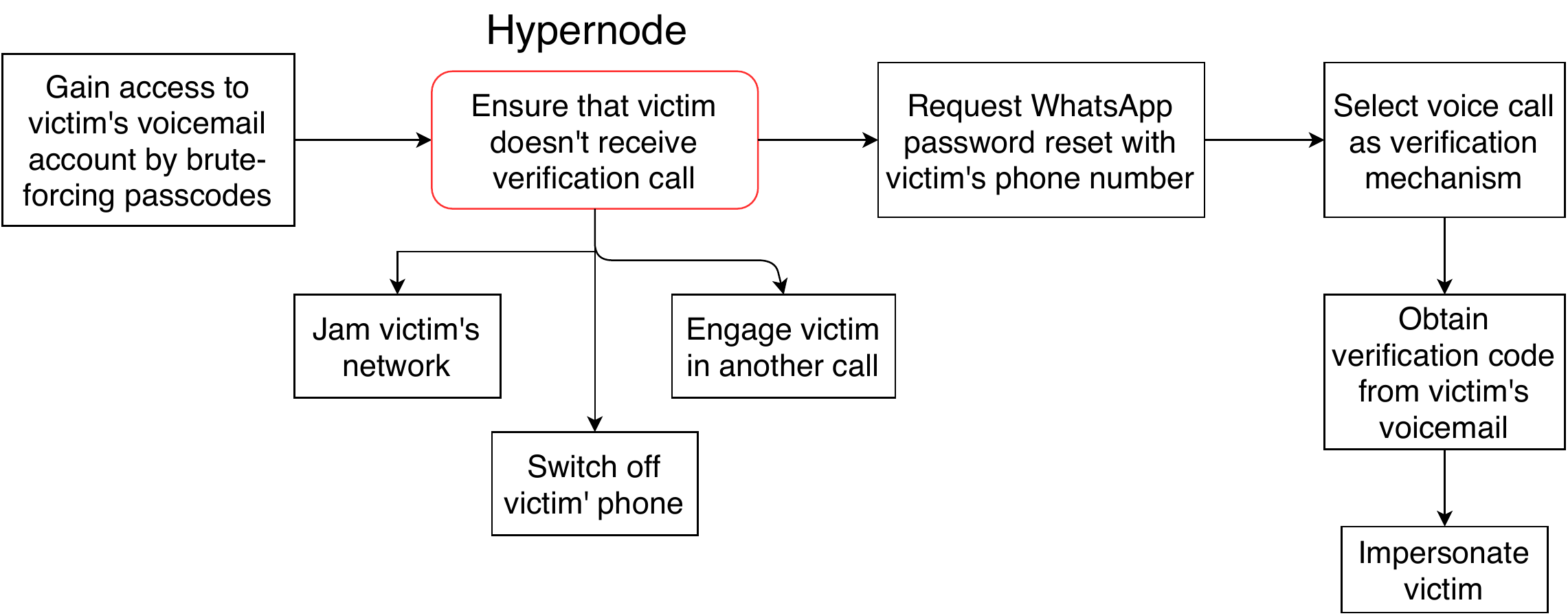}
    \caption{CFG of voicemail exploit}
    \label{fig:Voicemail}
    \end{figure*}
    
    \item \textbf{Insecurity of messages blocked in transit:} This security feature is intentionally disabled by WhatsApp as a trade-off for higher efficiency. The attack scenario involves Alice sending a message to Bob. Bob doesn't receive the message immediately because he is offline. In the meantime, the adversary Eve steals Bob's SIM card and uses it to verify herself as Bob on another device. Now, Eve receives the message from Alice which was meant for Bob~\cite{carpay2019whatsapp}. The CFG of this attack is shown in Fig.~\ref{fig:blocked}.
    \begin{figure*}[h!]
    \centering
    \includegraphics[scale=0.7]{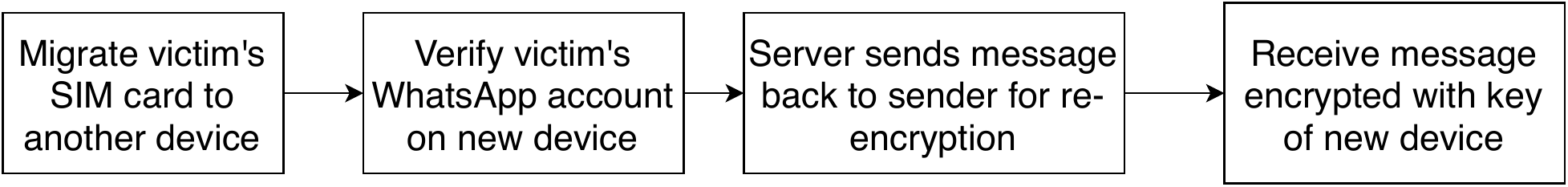}
    \caption{CFG of blocked messages in-transit exploit}
    \label{fig:blocked}
    \end{figure*}
    
\end{enumerate}

These attack vectors demonstrate that, although WhatsApp is highly secure as a stand-alone application, multiple implementation mechanisms on different platforms make it vulnerable to various attacks.

\subsubsection{Attacks on SDN framework}
\label{SDN_attacks}
The SDN stack in the 5G network architecture has introduced an entirely new attack surface with numerous threat vectors. The role of SDNs in virtualizing the network functions for greater flexibility and performance is pivotal for the realization of the promises made by 5G developers. Even at its nascent stage, it has already been adopted in multiple real-world systems like the Google datacenters.

Multiple network operating systems (NOS), also referred to as controllers, are available for the implementation of a SDN framework. The most popular amongst them are OpenDaylight~\cite{opendaylight}, Floodlight~\cite{floodlight} and POX~\cite{poxController}. All of them implement the most widely used OpenFlow~\cite{mckeown2008openflow} protocol for SDN communication. It has been shown that these NOSs are vulnerable to multiple classes of attacks~\cite{yoon2017flow}. A major concern is the absence of encryption of the messages exchanged between the controller and network switches.

We have considered the existing vulnerabilities of SDN and constructed an attack DAG that includes the CFGs of all such attacks. Then we implement cyber-analysis on this DAG to obtain new attacks on the SDN framework. The attacks that comprise the SDN attack DAG are:

\begin{enumerate}
    \item \textbf{Packet-In flooding:} The network hosts and switches send \textit{PacketIn} messages to the controller when it encounters a packet with unknown flow (a flow is the rule which instructs the switch where to forward the packet). The controller then sends an appropriate flow to the switch and the switch follows the specified flow rule in the future.
    
    If compromised switches send multiple PacketIn messages to the controller, the controller may exhaust all its resources in catering to them. This would lead the controller to an unpredictable state which may cause the network to fail. This is a classic DoS attack on the SDN controller. The CFG of this attack is shown in Fig.~\ref{fig:packet-In}.
    
    \begin{figure*}[h!]
    \centering
    \includegraphics[scale=0.7]{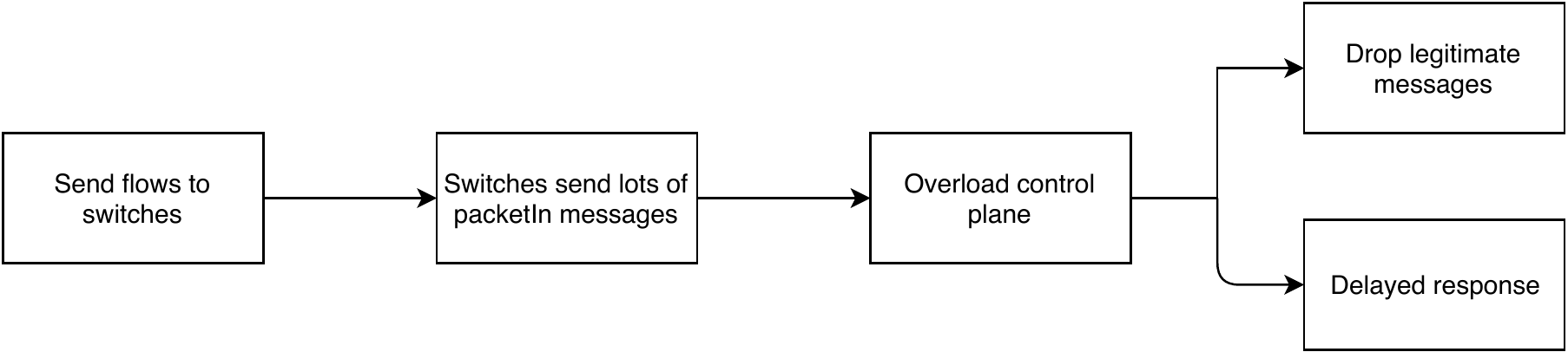}
    \caption{CFG of packet-In flooding attack on control plane of SDN}
    \label{fig:packet-In}
    \end{figure*}
    
    \item \textbf{Switch table flooding:} This is a DoS attack targetted towards exhausting the memory resources of the controller. The controller (or NOS) has a switch table which stores the information about all the switches present in the network. In the Floodlight controller, it is observed that modifying the \textit{features-reply} message of the SDN OpenFlow protocol with a unique datapacket ID (DPID) causes the controller to add a new entry to its switch table. Sending numerous maliciously-crafted messages can exhaust the switch table of the controller, causing the controller to disconnect the legitimate switches of the network. The CFG of this attack is shown in Fig.~\ref{fig:switch-table}.
    
    \begin{figure*}[h!]
    \centering
    \includegraphics[scale=0.7]{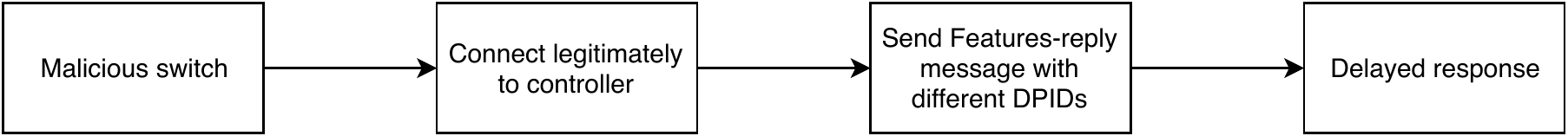}
    \caption{CFG of switch-table flooding attack on control plane of SDN}
    \label{fig:switch-table}
    \end{figure*}
    
    \item \textbf{Switch identification spoofing:} Due to weak authentication of the switches in the SDN protocol, a malicious switch may pose as a legitimate switch and request connection access with the controller. This causes the controller to disconnect the legitimate switch and connect with the malicious switch. The CFG of this attack is shown in Fig.~\ref{fig:switch-id_spoof}.
    
    \begin{figure*}[h!]
    \centering
    \includegraphics[scale=0.7]{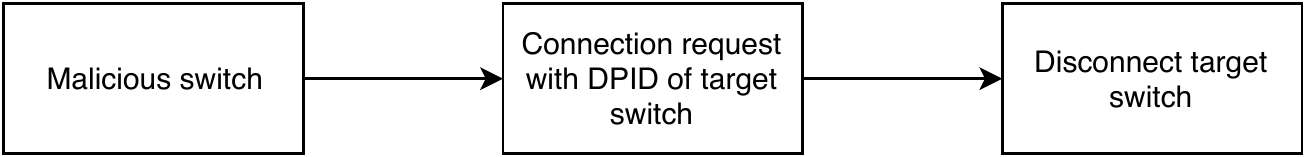}
    \caption{CFG of switch-id spoofing attack on SDN}
    \label{fig:switch-id_spoof}
    \end{figure*}
    
    \item \textbf{Malformed control message injection:} OpenFlow messages between the switches and the controller contains a header. An adversary can modify the contents of the header or the message itself to crash the SDN controller. In such a scenario, the controller might disconnect the switch. The CFG of this attack is shown in Fig.~\ref{fig:malformed_message}.
    
    \begin{figure*}[h!]
    \centering
    \includegraphics[scale=0.7]{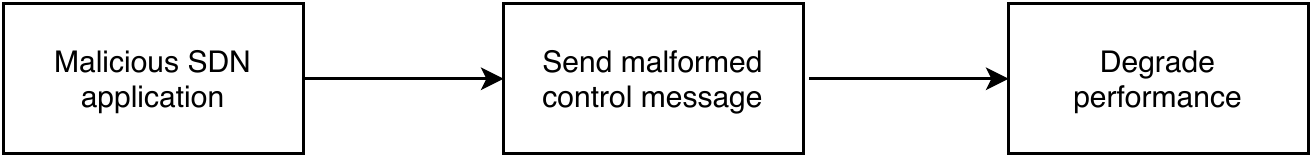}
    \caption{CFG of various malformed message attacks on SDN}
    \label{fig:malformed_message}
    \end{figure*}
    
    \item \textbf{System time manipulation:} Many SDN applications depend on the system time and other system variables. Modifying the system time may lead to disconnection of the switch which tries to perform time-sensitive operations. The CFG of this attack is shown in Fig.~\ref{fig:system_time}.
    
    \begin{figure*}[h!]
    \centering
    \includegraphics[scale=0.7]{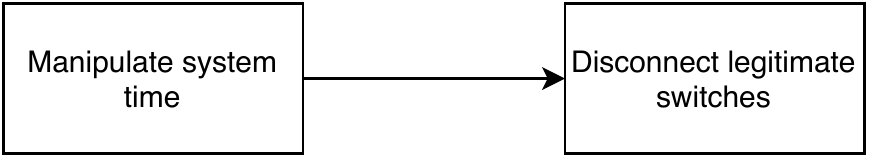}
    \caption{CFG of system time manipulation attack on SDN}
    \label{fig:system_time}
    \end{figure*}
    
    \item \textbf{Topology poisoning:} In order to achieve superior performance to traditional networks, an overall view of the entire network is available to the controller for efficient routing. The overall configuration of the network is implemented via host tracking services and link discovery services. These services can be exploited by multiple techniques (like node spoofing)to alter the topology of the network in the controller database. This lead to various attacks like eavesdropping attack and MiTM attack. The CFG of this attack is shown in Fig.~\ref{fig:topology_poisoning}.
    
    \begin{figure*}[h!]
    \centering
    \includegraphics[scale=0.7]{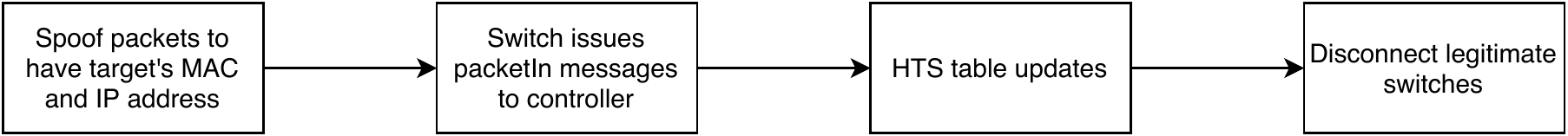}
    \caption{CFG of topology poisoning attack on SDN}
    \label{fig:topology_poisoning}
    \end{figure*}
    
    \item \textbf{Arbitrary system termination:} A malicious SDN application can call the system exit function. This will lead to the termination of the entire controller. All applications will be killed and all entries of the switch table will be lost. This shall be devastating for the data plane elements like the switches and hosts. The CFG of this attack is shown in Fig.~\ref{fig:arbitrary_termination}.
    
    \begin{figure*}[h!]
    \centering
    \includegraphics[scale=0.5]{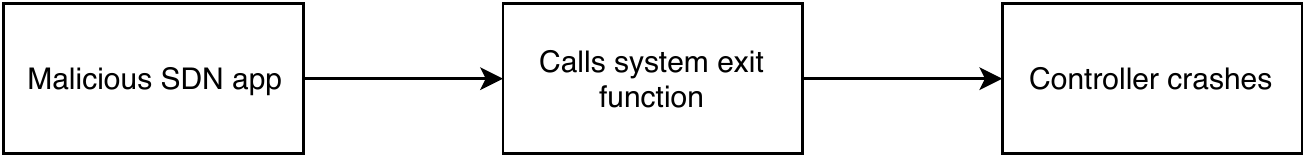}
    \caption{CFG of arbitrary system termination attack on SDN}
    \label{fig:arbitrary_termination}
    \end{figure*}
    
    \item \textbf{System resource exhaustion:} A malicious SDN application may utilize all of the resources of the host machine of the controller. This results in a DoS attack on the entire network. The CFG of this attack is shown in Fig.~\ref{fig:sys_resource_exhaustion}.
    
    \begin{figure*}[h!]
    \centering
    \includegraphics[scale=0.7]{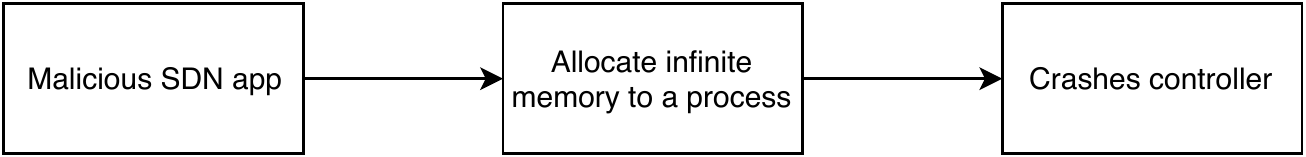}
    \caption{CFG of system resource exhaustion attack on SDN}
    \label{fig:sys_resource_exhaustion}
    \end{figure*}
    
    \item \textbf{Message delivery obstruction 1:} Lack of authorization of applications allows SDN malwares to alter the list of subscribers to packetIn messages. For example, App1 and App2 are subscribed to packetIn messages to the controller. A malware can alter the subscriber list by remove App2. This prevents App2 from receiving subsequent packetIn messages, thus hindering its proper functionality. The CFG of this attack is shown in ~\ref{fig:msg_delivery_obstruction}.
    
    \begin{figure*}[h!]
    \centering
    \includegraphics[scale=0.7]{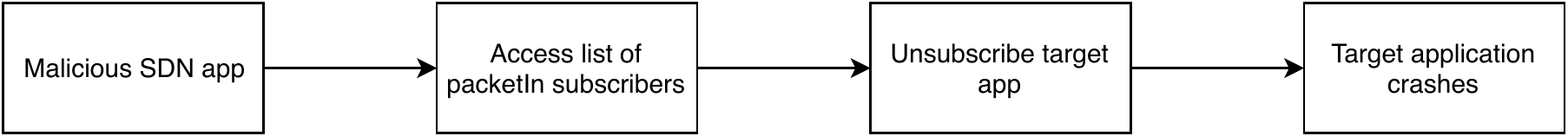}
    \caption{CFG of a message delivery obstruction attack on SDN}
    \label{fig:msg_delivery_obstruction}
    \end{figure*}
    
    \item \textbf{Message delivery obstruction 2:} Applications running on the SDN controller generally occurs sequentially in a chain. A malicious SDN application in the chain can drop a packet that was supposed to be forwarded to an application. This causes a misconfiguration in the network. The CFG of this attack is shown in ~\ref{fig:msg_delivery_obstruction2}.
    
    \begin{figure*}[h!]
    \centering
    \includegraphics[scale=0.7]{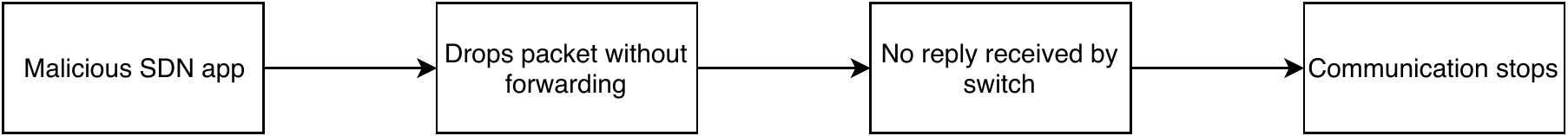}
    \caption{CFG of another message delivery obstruction attack on SDN}
    \label{fig:msg_delivery_obstruction2}
    \end{figure*}
    
    \item \textbf{Service chain jamming:} Similar to the previous attack, a malicious application can go into an infinite loop. This causes the subsequent applications in its control flow to be halted and the network comes to a standstill. The CFG of this attack is shown in ~\ref{fig:service_chain_jam}.
    
    \begin{figure*}[h!]
    \centering
    \includegraphics[scale=0.7]{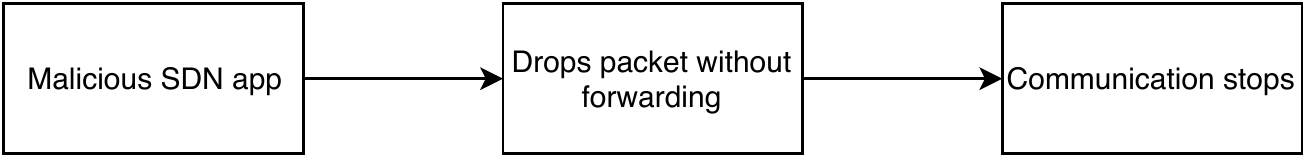}
    \caption{CFG of service chain jamming attack on SDN}
    \label{fig:service_chain_jam}
    \end{figure*}
    
    \item \textbf{Unauthorized application management:} SDN applications with maximal authorization provides maximal network programmability but introduces new attack vectors. Such applications can cause the termination of other applications. Such operations can cripple the network if a critical application like routing or packet forwarding is terminated. It may also produce various security threats if a security function like IDS or firewall is terminated. The CFG of this attack is shown in ~\ref{fig:unauthorized_app_management}.
    
    \begin{figure*}[h!]
    \centering
    \includegraphics[scale=0.7]{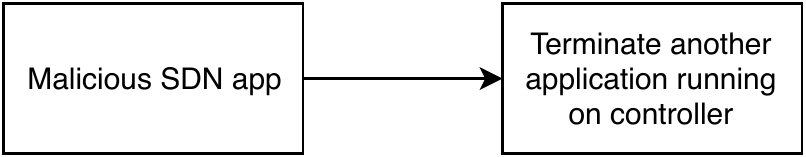}
    \caption{CFG of unauthorized application management attack on SDN}
    \label{fig:unauthorized_app_management}
    \end{figure*}
    
    \item \textbf{Flow rule modification:} The primary task of the controller is to generate flow rules. A malicious SDN application can modify the flow rules of a controller, leading to various attacks like MiTM and network failures. The CFG of this attack is shown in ~\ref{fig:flow_rule_modification}.
    
    \begin{figure*}[h!]
    \centering
    \includegraphics[scale=0.7]{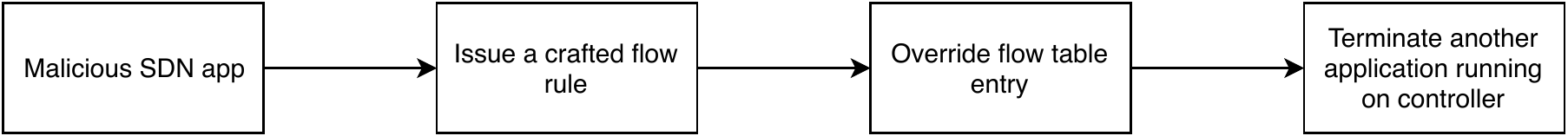}
    \caption{CFG of a flow rule modification attack on SDN}
    \label{fig:flow_rule_modification}
    \end{figure*}
    
    \item \textbf{Flow table flushing:} A malicious SDN application can flush all the flow table entries of the switch. This would cause the switch to issue a packetIn message for every incoming packet, thus leading to a degradation in network performance. The CFG of this attack is shown in ~\ref{fig:flow_table_flush}.
    
    \begin{figure*}[h!]
    \centering
    \includegraphics[scale=0.7]{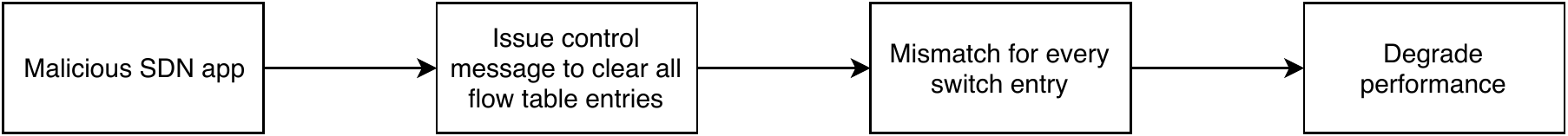}
    \caption{CFG of a flow table flush attack on SDN}
    \label{fig:flow_table_flush}
    \end{figure*}
    
    \item \textbf{Unauthorized network view manipulation:} An advantage of the SDN architecture is that it offers an overall view of the network topology. Most SDN controllers lack authorization of applications trying to modify internal storage. A SDN malware can alter the network topology stored in the controller, thus launching a topology poisoning attack. This affects all the other applications that make decisions based on this information. The CFG of this attack is shown in ~\ref{fig:unauth_nw_view}.
    
    \begin{figure*}[h!]
    \centering
    \includegraphics[scale=0.7]{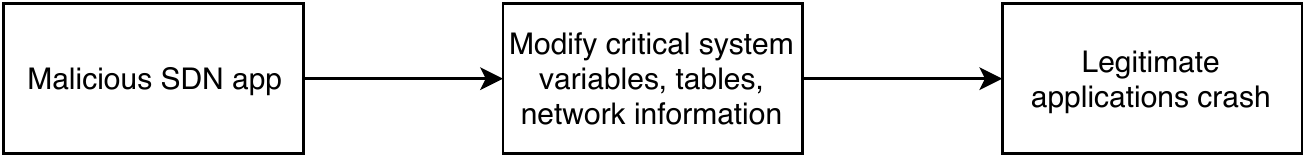}
    \caption{CFG of unauthorized network view manipulation attack on SDN}
    \label{fig:unauth_nw_view}
    \end{figure*}
    
    \item \textbf{Eavesdropping:} The control messages between the controller and the switches are unencrypted on most NOSs. Eavesdropping on the messages of the control channel can reveal information about the network topology.
    
    \item \textbf{MiTM on the control channel:} A MiTM attacker can actively modify the flow rules from "Forward" to "Drop" causing the switch to drop desired messages. The CFG of this attack is shown in ~\ref{fig:MiTM}.
    
    \begin{figure*}[h!]
    \centering
    \includegraphics[scale=0.7]{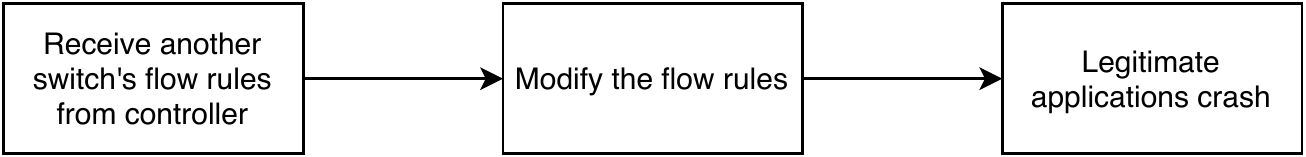}
    \caption{CFG of MiTM attack on control channel of SDN}
    \label{fig:MiTM}
    \end{figure*}
    
    \item \textbf{Flow-rule flooding:} A malicious SDN application can continuously generate flow rules till the memory of the switch is filled. Now, the switch cannot accept any new legitimate flow rules, thus failing the network. This is an instance of a DoS attack. The CFG of this attack is shown in ~\ref{fig:flow-rule_flooding}.
    
    \begin{figure*}[h!]
    \centering
    \includegraphics[scale=0.7]{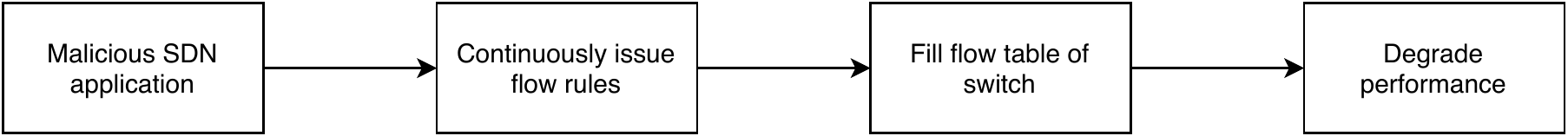}
    \caption{CFG of flow rule flooding attack on data plane of SDN}
    \label{fig:flow-rule_flooding}
    \end{figure*}
    
    \item \textbf{Switch firmware abuse:} Most OpenFlow switches run custom firmware implementation s with different capabilities. This spawns a new attack where the a particular flow may be forced to be processed in the software instead of the hardware by specifying the source and destination MAC addresses in the flow. Processing in software is much slower than hardware processing, thus leading to a degradation i network performance. The CFG of this attack is shown in ~\ref{fig:switch_firmware}.
    
    \begin{figure*}[h!]
    \centering
    \includegraphics[scale=0.7]{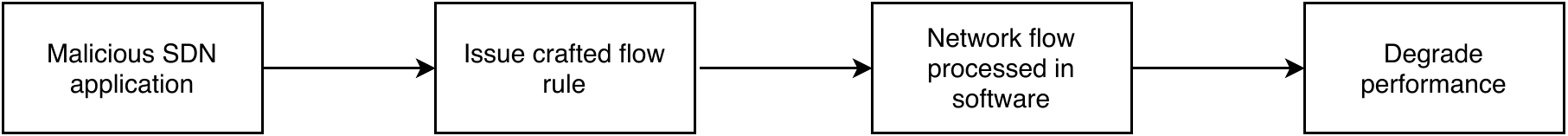}
    \caption{CFG of switch firmware abuse attack on SDN}
    \label{fig:switch_firmware}
    \end{figure*}
    
    \item \textbf{Malformed control message injection:} Manipulated control messages may be sent to the switches to cause the switch to end up in an unpredictable state, thus rendering it nonoperational. The CFG of this attack is shown in ~\ref{fig:malformed_msg}.
    
    \begin{figure*}[h!]
    \centering
    \includegraphics[scale=0.7]{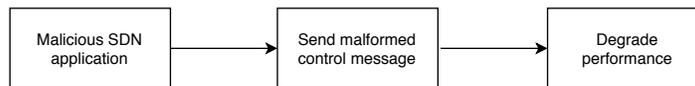}
    \caption{CFG of malformed message injection attack on SDN}
    \label{fig:malformed_msg}
    \end{figure*}
    
    \item \textbf{Data leakage:} SDN data plane elements are vulnerable to side-channel attacks like timing attacks. Such attacks can lead to fingerprinting attacks to compromise privacy. Side-channel attacks can also lead to loss of sensitive information like access control rules in the controller.
     
\end{enumerate}

\begin{figure*}
    \centering
    \includegraphics[scale=0.65]{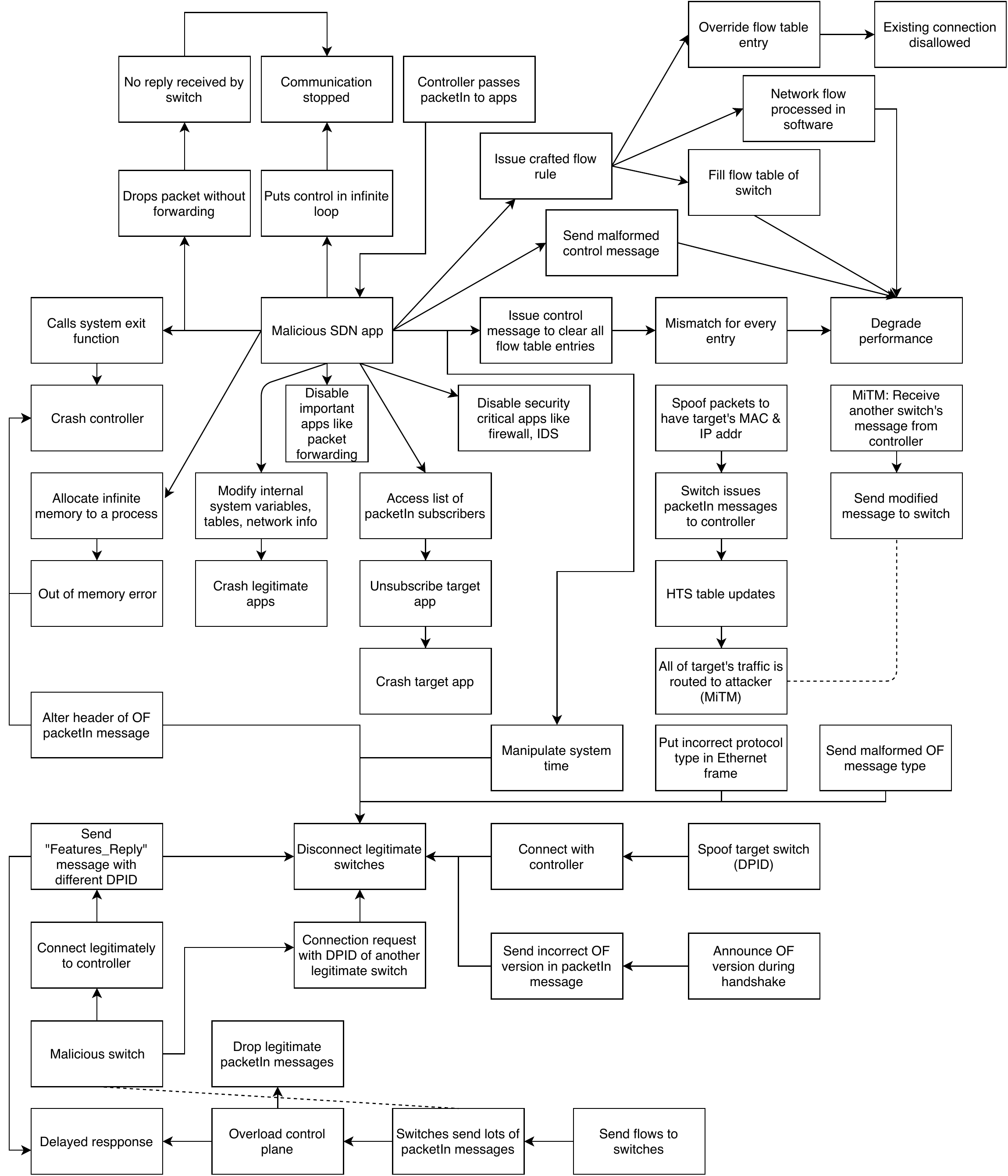}
    \caption{Attack DAG on SDN}
    \label{fig:SDN_DAG}
\end{figure*}

Combining all the CFGs mentioned above, we construct an attack CFG for SDN as shown in Fig.~\ref{fig:SDN_DAG}.

\subsection{Future Work}
Using ML on the system level operations has showed significant preliminary success~\cite{saha2022system, saha2021sharks}. We plan on extending this methodology to various other platforms and applications.
\subsubsection{Representation of attacks}
We will delve deep into each of the security categories mentioned above (along with other categories) and document all the possible attacks in each category. Then we decompose each attack into its constituent assembly-level instructions~\cite{saha2021sharks}. These assembly-level instructions are represented in the form of a CFG. We merge the CFGs of all the categories to form a unified DAG which encompasses all the known threats in the 5G ecosystem. Some of the detailed threats in 5G are documented in Tables~\ref{table:EPC_threats},~\ref{table:IMS_threats}. Similar tables can be constructed for RAN, SDN and NFV security.

\begin {table*}[h!]
\begin{center}
\caption {Mobile network security: Evolved Packet Core threats and challenges}
\begin{tabularx}{1.1\textwidth}{|c|p{3cm}|X|}
\hline
\textbf{Category} & \textbf{Threat} & \textbf{Description}\\
\hline
Availability & Flooding an interface & Attackers flood an interface resulting in DoS condition (DNS lookup, etc.) \\
\hline
Availability & Crashing a network element & Attackers crash a network element by sending malformed packets \\
\hline
Confidentiality & Eavesdropping & Attackers eavesdrop on sensitive data on control and user plane\\
\hline
Confidentiality & Data Leakage & Unauthorized access to sensitive data on the server\\
\hline
Integrity & Traffic modification & Attackers modify information during transit  (DNS redirection, etc.) \\
\hline
Integrity & Data modification & Attackers modify data on network element (change the network element configurations) \\
\hline
Control & Control the network & Attackers control the network via protocol or implementation flaw \\
\hline
Control & Compromise of network element & Attackers compromise of network element via management interface\\
\hline
Malicious insider & Insider attacks & Insiders make data modification on network elements, make unauthorized changes to network element configuration, etc. \\
\hline
Theft of service & Service free of charge & Attackers exploits a flaw to use services without being charged \\
\hline
\end{tabularx}
\label{table:EPC_threats}		
\end{center}
\end {table*}

\begin {table*}[h!]
\begin{center}
\caption {Threats and challenges in IP Multimedia Subsystems}
\begin{tabularx}{1.1\textwidth}{|c|p{4cm}|c|}
\hline
\textbf{Category} & \textbf{Threat} & \textbf{Description}\\
\hline
Availability & Flooding an interface & Distributed DoS (DDoS)/Telephony DoS (TDoS) via mobile end-points) \\
\hline
Availability & Crashing a network element & DoS/TDoS via rogue media streams and malformed packets  \\
\hline
Confidentiality & Eavesdropping & Eavesdropping via sniffing the SGi(Gm) interface\\
\hline
Confidentiality & Data Leakage & Unauthorized access to sensitive data on the
IMS Home Subscriber Server (HSS)\\
\hline
Integrity & Traffic modification & MiTM attack on SGi(Gm) interface \\
\hline
Integrity & Data modification & Session Initiation Protocol (SIP) messaging impersonation via spoofed SIP messages \\
\hline
Control & Control the network & Spam over Internet Telephony / unsolicited voice
calls resulting in Voice-SPAM/TDoS \\
\hline
Control & Compromise of network element & Compromise of network element via attacks from external IP networks\\
\hline
Malicious insider & Insider attacks & Malicious Insider makes unauthorized changes to IMSHSS configurations \\
\hline
Theft of service & Service free of charge & Theft of Service via SIP messaging impersonation \\
\hline
\end{tabularx}
\label{table:IMS_threats}		
\end{center}
\end {table*}

\subsubsection{Machine learning on DAG}
After the construction of the DAG, we characterize each node of the DAG with multiple binary-valued features. We train multiple ML models like SVM, K-nearest neighbors and decision trees with the node features. The ML model is trained to predict new possible branches in the DAG, giving rise to new attacks. Deep learning methods cannot be used in this scenario due to the insufficient data.

\subsubsection{Implementation on different frameworks}
This framework is a general approach which can be executed on any application. The algorithm remains the same but the DAG changes depending on the features and specifications of the framework. Our immediate focus is to implement this on a secure instant messaging application, like Signal or Whatsapp. We aim to test its effectiveness on different platforms like Android, iOS, Unix, etc. This analysis will help to comprehend the underlying dependencies of application layer security on the network and hardware stacks.

\subsubsection{Evaluation of the 5G stack}
5G network security is a challenging research problem as its infrastructure is entirely different from its predecessor. Many unknown vulnerabilities potentially exist across multiple levels in the hardware, software and network stacks of implementation. We aim to discover these weaknesses using our algorithm to make the 5G infrastructure safer and more secure. There are certain classes of attacks that can be easily evaded in 5G due to the separation of the data plane and the control plane. Multiple attack and defence measures are expected to be accelerated in the 5G ecosystem. We expect to get novel insights into the overall security framework of 5G by probing into these intricacies with ML.

\subsubsection{Reinforcement Learning for security}

Reinforcement learning (RL)~\cite{sutton1998introduction} can be utilized to provide real-time defence solutions. Similarly, it can also be utilized to come up with real-time attack vectors. We aim to use RL to study the implications of real-time attack scenarios in the 5G framework. A defensive or attacking agent can be deployed in a real-world scenario to learn the inconsistencies in the operations. Multi-agent RL may also prove to be useful in such scenarios. Counterfactual regret (CFR)~\cite{zinkevich2008regret} minimization games have shown better results than RL in certain classes of games (where the environment is not fully observable to the player) like poker. We would like to investigate the potential impact of CFR minimization games in 5G security.

\section{Evaluation Plan}
\subsection{Year 1}
\begin{itemize}
    \item Develop a unified framework for analyzing the interconnected security vulnerabilities between the hardware, software (microcode, kernel, OS and application layers) and network (including, cryptographic protocols, virtualization layers, and others) stacks with 5G network architectures.
    
    \item Study the dependencies of application layer security on the network layer security in the 5G framework.
    
    \item Analyze the security of cryptographic protocols and primitives proposed as part of the 5G framework.
    
    \item Analyze the security of a mobile application (which is secure at the application layer) running on 5G. This will give a deeper insight into the applicability and usability of the developed tool.
    
    \item Utilize deep learning techniques to analyze graphs representing various attacks and threat models. Recent graphical deep learning analysis tools like Graph2Vec ~\cite{narayanan2017graph2vec} enables us to use deep learning to analyze graphs. This has a potential of generating new methods of analyzing threat models.
    
\subsection{Year 2}
    \item Utilize the developed framework to analyze the security of multiple 5G-specific applications like autonomous vehicles, Internet-of-Things and smart healthcare.
    
    \item Extend the existing framework to other broader areas of security like application security, OS security, etc.
    
\subsection{Year 3}
    \item Integrate state-of-the-art methods of ML like RL to deploy real-time vulnerability detection systems.
    
    \item With the advent of 5G, the amount of data being transferred will increase manifold. We wish to study the implications of the magnanimous amount of data transfer on user privacy.
    
\end{itemize}

\bibliographystyle{IEEEtranN}
\bibliography{output}
\end{document}